\documentclass[aps,prc,floatfix,twocolumn,showpacs,amsmath]{revtex4}
\usepackage{dcolumn}
\usepackage{mathrsfs}
\usepackage{supertabular}
\usepackage{color,graphicx}
\usepackage[bookmarksopen]{hyperref}
\newcommand{\bm}{\bibitem}

\def  \a    {\alpha}

\def  \g    {\gamma}
\def  \G    {\Gamma}
\def  \l    {\lambda}
\def  \L    {\Lambda}

\def  \p    {\pi}

\def  \f    {\frac}
\def  \lt   {\left}
\def  \rt   {\right}

\def  \ra   {\rightarrow}

\def  \Dt   {\Delta}
\def  \ep   {\epsilon}
\def  \z    {\zeta}

\def  \be   {\begin{equation}}
\def  \ee   {\end{equation}}
\def  \ba   {\begin{array}}
\def  \ea   {\end{array}}
\def  \bea  {\begin{eqnarray}}
\def  \eea  {\end{eqnarray}}
\def  \nn   {\nonumber}
\def  \bd   {\begin{displaymath}}
\def  \ed   {\end{displaymath}}
\def  \bse  {\begin{subequations}}
\def  \ese  {\end{subequations}}
\def  \bwt  {\begin{widetext}}
\def  \ewt  {\end{widetext}}

\begin{document}
\title{$\pi$-$\eta$ mixing and charge symmetry violating 
NN potential in matter}
\author{Subhrajyoti Biswas}
\author{Pradip Roy}
\author{Abhee K. Dutt-Mazumder}
\address{Saha Institute of Nuclear Physics,
1/AF Bidhannagar, Kolkata-700 064, INDIA}

\medskip
\date{\today}
\begin{abstract}
We construct density dependent Class $III$ charge symmetry violating 
(CSV) potential due to mixing of $\pi$-$\eta$ meson with off-shell 
corrections. The density dependence enters through the non-vanishing 
$\pi$-$\eta$ mixing driven by both the neutron-proton mass difference 
and their asymmetric density distribution. The contribution of density
dependent CSV potential is found to be appreciably larger than the 
contribution of vacuum CSV potential.
\end{abstract}
\vspace{0.08 cm}
\maketitle

\section{introduction}

One of the interesting area of research in nuclear physics is the 
study of symmetries and their violation. The general goal of the 
research in this area is to find small but observable effects of 
charge symmetry violation (CSV) which might provide significant
insight into the dynamics of CSV interaction.

CSV of the $NN$ interaction refers to a difference between proton-
proton and neutron-neutron interactions only. It is most clearly
manifested in the $^1S_0$ scattering lengths {\em i.e.} the difference
between $pp$ and $nn$ scattering lengths at $^1$S$_0$ state is non-
zero \cite{Miller90,Howell98,Gonzalez99}. Other convincing evidence of
CSV comes from the binding energy difference of mirror nuclei which is
known as Okamoto-Nolen-Schifer (ONS) anomaly \cite{Nolen73,Okamoto64,
Garcia92}. The modern manifestation of CSV includes difference of 
neutron-proton form factors, hadronic correction to $g-2$ 
\cite{Miller06}, the observation of the decay of $\Psi^{\prime} (3686)
\ra (J/\Psi) \p^0 $ etc \cite{Miller06}.

The current understanding of CSV is that at the fundamental level, 
caused by the finite mass difference between  up $(u)$ and down $(d)$ 
quarks \cite{Miller90,Nolen69,Henley69,Henley79,Machleidt89,Miller95}.
As a consequence, at the hadronic level, charge symmetry (CS) is 
violated due to non-degenerate mass of nucleons.

There are various mechanisms that can lead to CSV in $NN$ interaction.
For example, neutral mesons with same spin parity with different 
isospin can mix at the fundamental level due to quark mass difference 
(at the hadronic level due to neutron-proton mass splitting). The most
important is the $\rho$-$\omega$ mixing, which according to Ref.
\cite{Henley79,McNamee75,Coon77,Blunden87,Sidney87,Machleidt01}
is claimed to be successful to explain CSV observables. The other 
examples are $\pi$-$\eta$ and $\pi$-$\eta^\prime$ mixing 
\cite{Coon87,Kim93,Piekarewicz93}. It is shown in Ref.\cite{Paul79} 
that $\pi$-$\eta^\prime$ is important as it is of opposite sign to 
$\pi$-$\eta$ mixing where individual contribution is known to be small.

In all the previous works, the mixing amplitude is taken to be either
constant or on-shall \cite{Blunden87,Sidney87}, which is not consistent 
for the construction of $NN$ potential as the mixing amplitude has strong
momentum dependence \cite{Cohen95,Piekarewicz92,Goldman92,Krein93,Connell94,
Coon97,Hatsuda94}. Even the $\rho$-$\omega$ mixing amplitude changes sign 
as one moves away from $\rho(\omega)$ pole to space-like region. It is important
to note that the mixing amplitude in the space-like region is relevant for the
construction of CSV potential.

Once the mixing amplitude is known, one can construct CSV potential 
by evaluating two body $NN$ scattering diagram involving mixed 
intermediate states like $\pi$-$\eta$ or $\rho$-$\omega$. It is to be 
noted that external legs can also contribute to the CSV if one 
incorporates relativistic corrections as has recently been shown in 
Ref.\cite{Biswas08}.

In matter, there can be another source of symmetry breaking if the 
ground state contains unequal number of neutrons ($n$) and protons 
($p$) giving rise to ground state induced mixing of various charged 
states like $\rho$-$\omega$, $\pi$-$\eta$ etc. meson even in the limit
$M_n=M_p$.

The matter induced mixing was studied in \cite{Abhee97,Broniowski98,
Abhee01,Kampfer04,Roy08,Biswas06}. But none of these deal with the 
construction of two-body potential except \cite{Saito03} where density
dependent CSV potential has been constructed considering only the 
effect of the scalar mean field on the nucleon mass excluding the 
possibility of matter driven mixing. Recently, the medium dependent 
CSV potential due to $\rho$-$\omega$ mixing has been constructed in
Ref.\cite{Biswas10}. It is also to be noted that such mixing 
amplitudes, in asymmetric nuclear matter (ANM), have non-zero 
contribution even if the quark or nucleon masses are taken to be equal
\cite{Abhee97,Broniowski98,Abhee01,Kampfer04,Roy08}. We consider both 
of these mechanisms to construct CSV $NN$ potential due to $\pi$-$\eta$ 
mixing. 

Although the vacuum contribution of $\pi$-$\eta$ mixing to CSV has 
been shown to be negligible as compared to $\rho$-$\omega$ mixing, we,
in this work, explore the role of medium dependent $\pi$-$\eta$ mixing
amplitude in constructing the CSV potential taking into account the 
contribution of external legs. 

\begin{figure}[htb]
\begin{center}
\resizebox{7.75cm}{2.75cm}{\includegraphics[]{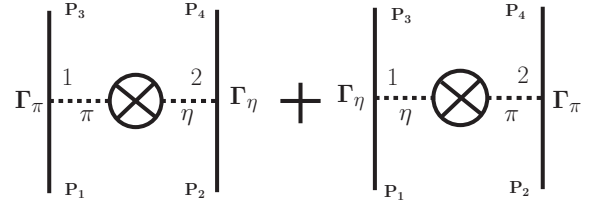}}
\caption{Feynman Diagrams that contribute to the construction 
of CSV NN potential in matter. Solid lines represent nucleons
and dashed lines stand for mesons. The crossed circles indicate
the symmetry breaking piece. \label{fig00}}
\end{center}
\end{figure}

Physically, in dense system, intermediate mesons might be absorbed and
re-emitted from the Fermi spheres. In symmetric nuclear matter (SNM) 
the emission and absorption  involving different isospin states like 
$\pi$ and $\eta$ cancel when the contributions of both the proton and 
neutron Fermi spheres are added provided the nucleon masses are taken 
to be equal. In ANM, on the other hand, the unbalanced contributions 
coming from the scattering of neutron and proton Fermi spheres, lead 
to the mixing which depends both on the density $(\rho_B)$ and the
asymmetry parameter $[\a = (\rho_n-\rho_p)/\rho_B] $. Inclusion
of this process is depicted by the third diagram in Fig.\ref{fig02}.

We present the formalism in Sec.II where the three momentum dependent
$\pi$-$\eta$ mixing amplitude is calculated to construct the CSV 
potential in matter. The numerical results are discussed in Sec.III. 
Finally, in Sec.IV, we summarise our results.

\section{formalism}

The following matrix element, which is required to construct the CSV
$NN$ potential is obtained from Fig.\ref{fig00}.

\nointerlineskip
\bea
\mathcal{M}^{NN}_{\pi\eta}(q)\!\! &=&\!\![\bar{u}_N(p_3)\tau_3(1)
\G_\pi(q)u_N(p_1)]\Dt_{\pi}(q)\Pi_{\pi\eta}(q^2) \nn \\
&\times &\!\!\Dt_{\eta}(q)[\bar{u}_N(p_4)\G_{\eta}(-q)u_N(p_2)]\nn \\
&+&\!\![\bar{u}_N(p_3)\G_\eta(q) u_N(p_1)]\Dt_{\eta}(q)\Pi_{\pi\eta}(q^2)\nn \\
&\times &\!\!\Dt_{\pi}(q)[\bar{u}_N(p_4)\tau_3(2)\G_{\pi}(-q) u_N(p_2)].
\label{ma0}
\eea

Here $u_N$'s represent Dirac spinors, $\Pi_{\pi\eta}(q^2)$ is the
$\pi$-$\eta$ mixing amplitude, $p_i,~(i=1-4)$ and $q$ are the four 
momenta of nucleon and meson, respectively. $\tau_3(1)$  and 
$\tau_3(2)$ are isospin operators at vertices `$1$' and `$2$' 
({\em see} Fig.\ref{fig00}). The vertex factor is denoted by $\G_j(q),
~(j=\pi,\eta)$ and $\Dt_j(q)$ stands for meson propagator given by

\bea
\Dt^{-1}_j(q^2)=q^2-m^2_j. \label{meson:prop}
\eea

In the limit $q_0 \ra 0$, Eq.(\ref{ma0}) gives the momentum 
space CSV $NN$ potential, $V^{NN}_{CSV}({\bf q})$. 

\begin{figure}[htb]
\begin{center}
\resizebox{7.5cm}{1.35cm}{\includegraphics[]{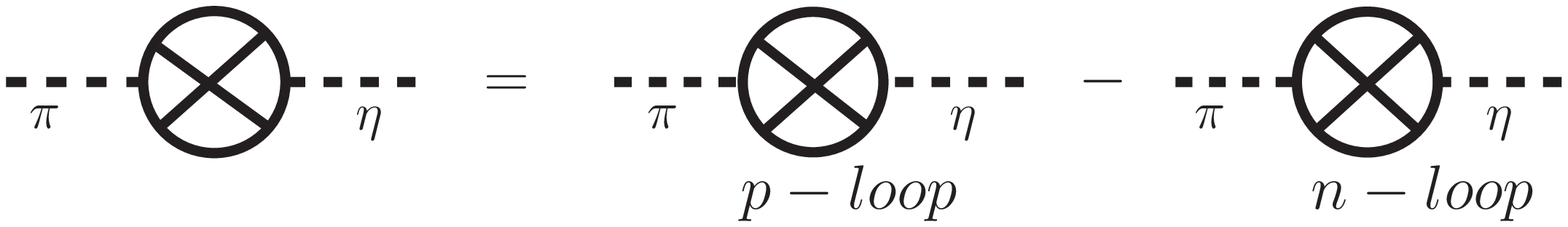}}
\caption{The mixing amplitude is generated by the difference 
between proton and neutron loops. \label{fig01}}
\end{center}
\end{figure}

In the present calculation mixing is assumed to be generated 
by the $N\bar{N}$ loops and the mixing amplitude $\Pi_{\pi\eta}(q^2)$ 
is generated by the difference between proton and neutron loop 
contributions as shown in Fig.\ref{fig01}:

\bea
\Pi_{\pi\eta}(q^2) = \Pi^{(p)}_{\pi\eta}(q^2)-\Pi^{(n)}_{\pi\eta}(q^2),
\label{mix_amp}
\eea

\noindent where $\Pi^{(p)}_{\pi\eta}(q^2)$ or $\Pi^{(n)}_{\pi\eta}
(q^2)$ is the $\pi$-$\eta$ mixing self-energy. The origin of the
relative sign between proton and neutron loops in Eq.(\ref{mix_amp}) 
is due to the different signs involved in the coupling of $\pi^0$ and 
$\eta$ to proton and neutron. The one loop contribution  to the mixing
self-energy is given by

\nointerlineskip
\bea
&\!\!&i\Pi^{(N)}_{\pi\eta}(q^2)\!\! =\!\!\!\int \!\! \f{d^4k}{(2\pi)^4}
{\rm Tr}\!\lt[\!\G_\pi (q) G_N(k) \G_\eta(-q) G_N(k\!+\!q)\!\rt],\nn\\
&\!\!&\label{se}
\eea

\noindent where the subscript $N$ stands for nucleon index 
({\em i.e.} $N=p$ or $n$), $k=(k_0,{\bf k})$ denotes the four 
momentum of the nucleon in the loops. The main ingredient of our 
calculation is the in-medium nucleon propagator $G_N$ which consists
of a free $(G^F_N)$ and a density dependent $(G^D_N)$ parts 
\cite{Serot86}:

\bse
\label{nucl:prop}
\bea
G^F_N(k)\!\!&=&\!\!\frac{k\!\!\!/+M_N}{k^2-M^2_N+i\z}, 
\label{nucl:prop_vac} \\
G^D_N(k)\!\!&=&\!\!\frac{i\pi}{E_N}(k\!\!\!/+M_N)\delta(k_0-E_N)
\theta(k_N-|{\bf k}|).
\label{nucl:prop_med}
\eea
\ese

\noindent $E_N=\sqrt{M^2_N+k^2_N}$ is the nucleon energy where $k_N$
and $M_N$ denote Fermi momentum and mass of the nucleon, respectively. 
Note that delta function in Eq.(\ref{nucl:prop_med}) indicates the 
nucleons are on-shell while $\theta(k_N-|{\bf k}|)$ ensures that 
propagating nucleons have momentum less than $k_N$.

\begin{figure}[htb]
\begin{center}
\resizebox{7.5cm}{1.75cm}{\includegraphics[]{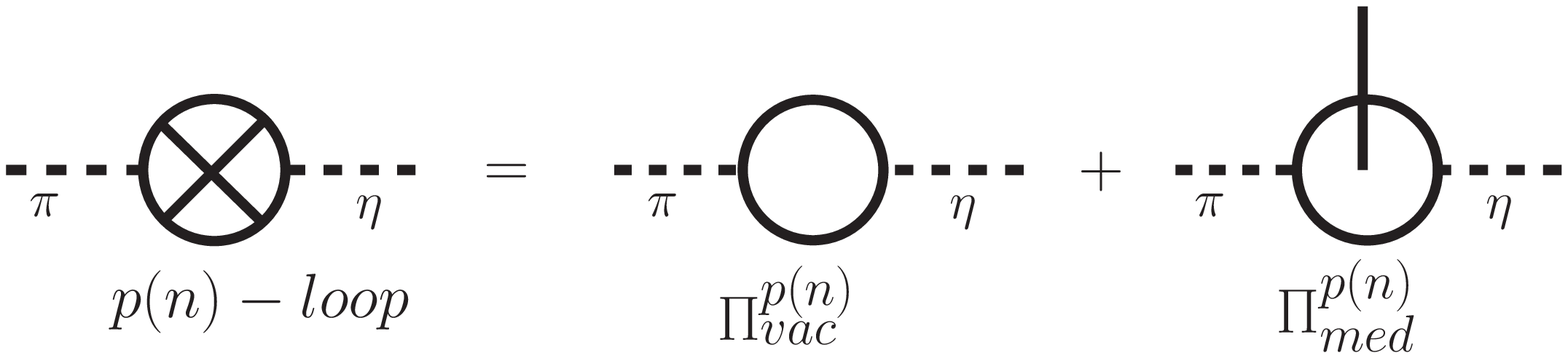}}
\caption{The mixing self-energy contains a vacuum part and 
a medium part. 
\label{fig02}}
\end{center}
\end{figure}

Likewise the in-medium nucleon propagator $G_N$ the mixing 
self-energy $\Pi^{(N)}_{\pi\eta}(q^2)$ contains a vacuum 
$[\Pi^{(N)}_{vac}(q^2)]$ and a density dependent $[\Pi^{(N)}_{med}
(q^2)]$ parts as shown in Fig.\ref{fig02}. It is to be noted that 
the density dependent part is given by the combination of $G^F_NG^D_N
+G^D_NG^F_N$ corresponds to scattering that we have discussed already,
whereas the term proportional to $G^D_NG^D_N$ vanishes for low energy 
excitation \cite{Chin77}. The vacuum part, {\em viz.} 
$[\Pi^{(N)}_{vac}(q^2)]$ on the other hand involves $G_N^FG_N^F$ 
which gives rise to usual CSV part of the potential due to the 
splitting of the neutron and proton mass.

The vacuum mixing contribution of CSV $NN$ potential can be used 
to calculate the difference between $nn$ and $pp$ scattering lengths, 
$\Dt a = a_{pp}-a_{nn}$, at $^1S_0$ state \cite{Biswas08}.

\subsection{Pseudoscalar coupling}

First we consider pseudoscalar (PS) coupling of nucleons to the mesons
to describe $\pi NN$ and $\eta NN$ interactions which are represented 
by the following effective Lagrangians:

\bse
\bea
\mathcal{L}^{PS}_{\pi NN} &=&-i{\rm g_\pi}\bar{\Psi}\g_5{\bf \tau}
\cdot{\bf \Phi}_\pi \Psi,
\label{ps:lag_pi}\\
\mathcal{L}^{PS}_{\eta NN} &=&-i{\rm g_\eta}\bar{\Psi}\g_5\Phi_\eta\Psi,
\label{ps:lag_eta}
\eea
\ese

where $\Psi$ and $\Phi$ represent the nucleon and meson fields,
respectively, and ${\rm g}_j$'s stand for the meson-nucleon coupling 
constants. For PS coupling the vertex factor $\G_j=-i{\rm g}_j\g_5, 
~(j=\pi,\eta)$.

Now we proceed to calculate the CSV $NN$ potential. For this purpose
we need to calculate the $\pi$-$\eta$ mixing self-energy using 
Eq.(\ref{se}). First we consider mixing in vacuum. After performing 
trace calculation, the vacuum contribution of $\pi$-$\eta$ mixing 
self-energy is found to be

\bea
\Pi^{(N)}_{vac}(q^2)\!\!&=&\!\!4i{\rm g_\pi g_\eta}
\!\!\int\!\!\f{d^4k}{(2\pi)^4}\!\lt[\!\f{M^2_N\!-
\!k\cdot(k\!+\!q)}{(k^2-M^2_N)((k+q)^2-M^2_N)}\!\rt]. \nn \\
&&\label{ps:vac_se_00}
\eea

From the dimensional counting it is found that the integral
of Eq.(\ref{ps:vac_se_00}) is divergent. We use dimensional 
\cite{Hooft73,Peskin95,Cheng06} regularization to isolate the
singularities in Eq.(\ref{ps:vac_se_00}) which reduces to
\cite{Piekarewicz93}

\bea
\Pi^{(N)}_{vac}(q^2)\!\!&=&\!\!\f{{\rm g_\pi g_\eta}}{4\pi^2}\!\!
\lt[\!\f{q^2}{3}\!+\!\!\lt(\!M^2_N\!-\!\f{q^2}{2}\!\rt)\!\lt(\!1\!
+\!\f{1}{\ep}\!-\!\g_E\!+\!\ln(4\pi\mu^2)\!\rt)\rt.\nn\\
&-&\!\!\lt.\!\!\int^1_0\!\!\!dx [M^2_N\!-\!3q^2x(1\!-\!x)]
\ln(M^2_N\!-\!q^2x(1\!-\!x))\!\rt].\nn \\
&&\label{ps:vac_se_01}
\eea   

In Eq.(\ref{ps:vac_se_01}) $\mu$ is an arbitrary scale parameter, 
$\g_E$ is the Euler-Mascheroni constant and $\ep = 2-D/2$, where $D$ 
stands for the dimension of the integral. Notice, $\ep$ in 
Eq.(\ref{ps:vac_se_01}) contains the singularity and it diverges 
as $D \ra 4$. The divergences of Eq.(\ref{ps:vac_se_01}) can be 
removed by adding appropriate counterterms \cite{Matsui82}.

It is clear from Eq.(\ref{ps:vac_se_01}) that unlike $\rho$-$\omega$ 
mixing amplitude, the singularities cannot be removed by simply 
subtracting the neutron loop contribution from the proton loop 
contribution. This is because of the singular term proportional 
to the mass term. But one can eliminate this singular term by 
subtracting $\Pi^{(N)}_{vac}(q^2=0)$ from $\Pi^{(N)}_{vac}(q^2)$ 
which is  

\bea
\hat{\Pi}^{(N)}_{vac}(q^2)\!\!&=&\!\!\Pi^{(N)}_{vac}(q^2)\!
-\!\Pi^{(N)}_{vac}(q^2\!=\!0)\nn \\
&=&\!\!\f{{\rm g_\pi g_\eta}}{4\pi^2}
\!\lt[\!\f{q^2}{3}\!+\!M^2_N\ln M^2_N \rt.\nn \\
&-&\lt.\!\f{q^2}{2}\!\lt(\!1+\!\f{1}{\ep}\!-\!\g_E\!+
\!\ln(4\pi\mu^2)\!\rt)\rt.\nn\\
&-&\!\!\lt.\!\!\int^1_0\!\!dx [M^2_N\!-\!3q^2x(1\!-\!x)]
\ln\lt(\!M^2_N\!-\!q^2x(1\!-\!x)\!\rt)\!\rt].\nn \\
&& \label{ps:vac_se_02}
\eea

Note that $\hat{\Pi}^{(N)}_{vac}(q^2)$ is not finite but the divergent
part proportional to the mass term has been removed. Now one can 
easily obtain finite $\pi$-$\eta$ mixing amplitude in vacuum by 
subtracting $\hat{\Pi}^{(n)}_{vac}(q^2)$ from $\hat{\Pi}^{(p)}_{vac}
(q^2)$. 

\bea
\Pi^{PS}_{vac}(q^2)\!\!&=&\!\f{{\rm g_\pi g_\eta}}{4\pi^2}\!
\lt[\!q^2\ln\lt(\f{M_p}{M_n}\rt)\rt. \nn \\
&+&\!\!\lt.q\sqrt{4M^2_p\!-\!q^2}\tan^{-1}\lt(\f{q}
{\sqrt{4M^2_p\!-\!q^2}}\rt)\rt.\nn\\
&-&\!\!\lt.q\sqrt{4M^2_n\!-\!q^2}\tan^{-1}\lt(\f{q}
{\sqrt{4M^2_n\!-\!q^2}}\rt)\rt].
\label{ps:vac_se_03}
\eea

Eq.(\ref{ps:vac_se_03}) is $q^2$ dependent $\pi$-$\eta$ mixing
amplitude. We obtain $\Pi^{PS}_{vac}(q^2=m^2_\eta) = -1197$ MeV$^{-2}$, 
while experimentally it is found that $\Pi^{PS}_{vac}(q^2=m^2_\eta)=
-4200$ MeV$^{-2}$ \cite{Piekarewicz93}. In this equation we substitute
 $q_0 =0$ to obtain three momentum dependent mixing amplitude 
$\Pi_{vac}({\bf q}^2)$ which is required for the construction of CSV
potential. Now expanding the mixing amplitude $\Pi^{PS}_{vac}({\bf q}^2)$
in terms of ${\bf q}^2/{M^2_N}$  and keeping the lowest order we obtain

\bea
\Pi^{PS}_{vac}({\bf q}^2)&=& -a_1{\bf q}^2, 
\label{ps:vac_amp}
\eea 

\noindent where $a_1=\f{{\rm g_\pi g_\eta}}{4\pi^2}\ln\lt(\f{M_p}{M_n}\rt)$ 
and if $M_p=M_n$, mixing amplitude in vacuum {\em i.e.} $\Pi^{PS}_{vac}
({\bf q}^2)$ vanishes. This implies that CSV $NN$ potential in vacuum 
does not exist for $M_p=M_n$.

Now we calculate the density dependent part of the $\pi$-$\eta$
mixing self-energy which is denoted by $\Pi^{(N)}_{med}(q^2)$. 
After performing trace calculation and $k_0$ integration it, 
reads 

\bea
\Pi^{(N)}_{med}(q^2)\!\!&=&\!\!\!-8{\rm g_\pi g_\eta}
\!\!\int^1_0\!\!\!\f{d^3{\bf k}}{(2\pi)^3E_N}\!
\lt[\!\f{(k\cdot q)^2}{q^4\!-\!4(k\cdot q)^2}\!\rt]
\!\theta(k_N\!-\!|{\bf k}|). \nn \\
&& \label{ps:med_se_00}
\eea  

The above equation with the substitution $E_N\simeq M_N$ and $q_0=0$ 
yields 

\bea
\Pi^{(N)}_{med}({\bf q}^2)\!\!&=&\!\!\f{{\rm g_\pi g_\eta}}{\pi^2M_N}
\lt[\f{k^3_N}{3}-\f{{\bf q}^2k_N}{8}\rt. \nn \\
&-&\lt.\!\f{{\bf q}}{8}\!\lt(\!k^2_N\!-\!\f{{\bf q}^2}{4}\!\rt)\!
\ln\lt(\!\f{{\bf q}\!+\!2k_N}{{\bf q}\!-\!2k_N}\!\rt)\rt]. 
\label{ps:med_se_01}
\eea

Eq.(\ref{ps:med_se_01}) represents three momentum dependent medium
part of the $\pi$-$\eta$ mixing self-energy. The mixing amplitude, 
as mentioned earlier, is again generated by the difference between 
contributions from the proton and neutron loops. This medium part 
of the $\pi$-$\eta$ mixing amplitude, after suitable expansion 
in terms of $\f{{\bf q}}{k_N}$ reads

\bea
\Pi^{PS}_{med}({\bf q}^2)\!\!&=&\!\!a^\prime_1-b^\prime_1{\bf q}^2, 
\label{ps:med_amp} 
\eea  

\noindent where the leading order contribution has been considered and 

\bse
\bea
a^\prime_1 &=& \f{{\rm g_\pi g_\eta}}{3\pi^2}
\lt(\f{k^3_p}{M_p}\!-\!\f{k^3_n}{M_n}\rt),\\
b^\prime_1 &=& \f{{\rm g_\pi g_\eta}}{4\pi^2}
\lt(\f{k_p}{M_p}\!-\!\f{k_n}{M_n}\rt).
\eea
\ese

Following Eq.(\ref{ma0}) and considering the contributions 
of external nucleon legs one obtains the momentum space potential 
given as

\bea
V^{NN}_{CSV}({\bf q}^2)&=&T^+_3\f{{\rm g_\pi g_\eta}}{4M^2_N}
({\bf \sigma_1}\cdot{\bf q})({\bf \sigma_2}\cdot{\bf q})\nn \\
&\times& \f{\Pi^{PS}_{\pi\eta}({\bf q}^2)}{({\bf q}^2+m^2_\pi)
({\bf q}^2+m^2_\eta)} \nn \\
&\times& \lt[1-\f{{\bf q}^2}{8M^2_N}-\f{{\bf P}^2}{2M^2_N} \rt], 
\label{ps:pot_mom}
\eea

\noindent where $T^+_3=\tau_3(1)+\tau_3(2)$ and ${\bf P}=
({\bf p}_1+{\bf p}_3)/2 = ({\bf p}_2+{\bf p}_4)/2$. Eq.(\ref{ps:pot_mom}) 
presents CSV class $II$ potential in momentum space. In contrast to 
$\pi$-$\eta$ mixing, $\rho$-$\omega$ mixing produces both class $III$
and class $IV$ $NN$ interactions \cite{Miller06,Piekarewicz92,Piekarewicz93}.
Note that the terms within the square bracket in Eq.(\ref{ps:pot_mom}) 
are the contributions from the external legs. 

The coordinate space CSV $NN$ potential is obtained by Fourier 
transformation of Eq.(\ref{ps:pot_mom}).

\bea
V^{NN}_{CSV}(r)&=& V^{NN}_{vac}(r)+V^{NN}_{med}(r),
\eea

where $V^{NN}_{vac}(r)$ represents CSV $NN$ potential in vacuum and
$V^{NN}_{med}(r)$ is the CSV $NN$ potential due to density driven 
mixing. Explicitly $V^{NN}_{vac}(r)$ and $V^{NN}_{med}(r)$ are given 
by   

\bse
\bea
V^{NN}_{vac}(r)\!\!&=&\!\!-T^+_3\f{{\rm g_\pi g_\eta}a_1}{48\pi M^2_N}
\lt[\!\f{m^5_\pi U(x_\pi)\!-\!m^5_\eta U(x_\eta)}{m^2_\eta-m^2_\pi}\!\rt], 
\label{ps:pot_cor_vac_noff} \\
V^{NN}_{med}(r)\!\!&=&\!\!-T^+_3\f{{\rm g_\pi g_\eta}}{48\pi M^2_N}
\lt[\!a^\prime_1\lt(\f{m^3_\pi U(x_\pi)\!-\!m^3_\eta U(x_\eta)}
{m^2_\eta-m^2_\pi}\rt)\rt.\nn \\
\!\!&+&\!\!\lt.\lt(\!\f{a^\prime_1}{8M^2_N}\!+\!b^\prime_1\!\rt)
\lt(\f{m^5_\pi U(x_\pi)\!-\!m^5_\eta U(x_\eta)}{m^2_\eta-m^2_\pi}\rt)\!\rt].
\label{ps:pot_cor_med_noff}
\eea
\ese

\noindent Here

\bse
\bea
U(x_i)&=&Y_0(x_i)({\bf \sigma_1}\cdot{\bf \sigma_2})
+S_{12}(\hat{{\bf r}})Y_2(x_i)\\
Y_2(x_i)&=&\lt(1+\f{3}{x_i}+\f{3}{x^2_i}\rt)Y_0(x_i)\\
S_{12}(\hat{{\bf r}})&=&3({\bf \sigma_1}\cdot\hat{{\bf r}})
({\bf \sigma_2}\cdot\hat{{\bf r}})
-({\bf \sigma_1}\cdot{\bf \sigma_2})
\eea
\ese

\noindent where $x_i=m_ir$, $i=\pi,\eta$ and $Y_0(x_i)
=\f{e^{-x_i}}{x_i}$.\\ 

Since mesons and nucleons are not point particles and they have
internal structures one needs to incorporate vertex corrections which, 
in principle, can be calculated using renormalizable models based on 
hadronic degrees of freedom. In the present calculation following 
phenomenological form factors have been used to incorporate the vertex 
corrections,

\bea
F_i({\bf q^2})&=&\lt(\f{\Lambda^2_i-m^2_i}{\Lambda^2_i+{\bf q}^2}\rt),
~~i=\pi,\eta.
\eea

Here $\Lambda_i$ is the cut-off parameter. With the inclusion of 
form factors Eqs.(\ref{ps:pot_cor_vac_noff}) and (\ref{ps:pot_cor_med_noff})
reduce to

\bea
V^{NN}_{vac}(r)\!\!&=&\!\!-T^+_3\f{{\rm g_\pi g_\eta}a_1}{48\pi M^2_N}
\lt[\!\lt(\f{a_\pi m^5_\pi U(x_\pi)\!-\!a_\eta m^5_\eta U(x_\eta)}
{m^2_\eta-m^2_\pi}\rt)\rt.\nn\\
\!\!&-&\!\!\lambda\!\lt.\lt(\f{b_\pi m^5_\pi U(X_\pi)\!-
\!b_\eta m^5_\eta U(X_\eta)}{m^2_\eta-m^2_\pi}\rt) \rt], 
\label{ps:pot_cor_vac_wff}
\eea

and

\bea
V^{NN}_{med}(r)\!\!&=&\!\!-T^+_3\f{{\rm g_\pi g_\eta}}{48\pi M^2_N}
\lt[\!a^\prime_1\lt(\f{a_\pi m^3_\pi U(x_\pi)\!-
\!a_\eta m^3_\eta U(x_\eta)}{m^2_\eta-m^2_\pi}\rt)\rt.\nn\\
\!\!&+&\!\!\lt.\!\lt(\f{a^\prime_1}{8M^2_N}+b^\prime_1\rt)
\lt(\f{a_\pi m^5_\pi U(x_\pi)\!-\!a_\eta m^5_\eta U(x_\eta)}
{m^2_\eta-m^2_\pi}\rt)\rt.\nn\\
\!\!&-&\!\!\lambda\!\lt.\lt\{\!a^\prime_1
\lt(\!\f{b_\pi m^3_\pi U(X_\pi)\!-\!b_\eta m^3_\eta U(X_\eta)}
{m^2_\eta-m^2_\pi}\!\rt)\rt.\rt. \nn \\
\!\!&+&\!\!\lt.\lt.\!\lt(\!\f{a^\prime_1}{8M^2_N}\!+\!b^\prime_1\!\rt)\!
\lt(\!\f{b_\pi m^5_\pi U(X_\pi)\!-\!b_\eta m^5_\eta U(X_\eta)}
{m^2_\eta-m^2_\pi}\!\rt)\!\rt\}\! \rt], 
\label{ps:pot_cor_vac_wff}
\eea

\noindent where $X_i=\Lambda_i r$ and 

\bse
\bea
a_i&=& \lt(\f{\L^2_j-m^2_j}{\L^2_j-m^2_i}\rt),\\
b_i&=& \lt(\f{\L^2_j-m^2_j}{m^2_j-\L^2_i}\rt),~~i \neq j,\\
\l&=& \lt(\f{m^2_\eta-m^2_\pi}{\L^2_\eta-\L^2_\pi}\rt).
\eea
\ese

\subsection{Pseudovector coupling}

Now we consider pseudovector (PV) coupling of nucleons to the mesons
to describe $\pi NN$ and $\eta NN$ interactions which are represented 
by the following effective Lagrangians:

\bse
\bea
\mathcal{L}^{PV}_{\pi NN} & = & -\f{{\rm g_\pi}}{2M_N}\bar{\Psi}
\g_5\g^\mu\partial_\mu{\bf \tau}\cdot{\bf \Phi}_\pi \Psi,
\label{pv:lag_pi}\\
\mathcal{L}^{PV}_{\eta NN} & = & -\f{{\rm g_\eta}}{2M_N}\bar{\Psi}
\g_5\g^\mu\partial_\mu\Phi_\eta \Psi,
\label{pv:lag_eta}
\eea
\ese

where $\Psi$, ${\bf \Phi}$, ${\bf \tau}$ and ${\rm g}$'s have already
been defined in the previous subsection, and the vertex factors 
$\G_j=i{\rm g}_j\g_5\g^\mu q_\mu,~(j=\pi,\eta)$. The mixing self-energy 
in vacuum is given by

\bea
\Pi^{(N)}_{vac}(q^2)\!\!&=&\!\!4i\lt(\f{{\rm g_\pi}}{2M_N}\rt)
\lt(\f{{\rm g_\eta}}{2M_N}\rt) \!\!\int\!\!\f{d^4k}{(2\pi)^4}\nn \\
\!\!&\times&\!\!\lt[\!\f{q^2(M^2_N\!-\!k\cdot(k\!+\!q))\!
-\!2q\cdot(k\!+\!q)(k\cdot q)}{(k^2\!-\!M^2_N)
((k\!+\!q)^2\!-\!M^2_N)}\!\rt].\nn \\
&& \label{pv:vac_se_00}
\eea

Note that the above integral is also divergent divergent 
and we use dimensional regularization to isolate singularities 
which reduces to

\bea
\Pi^{(N)}_{vac}(q^2)\!\!&=&\!\!\f{{\rm g_\pi g_\eta}}{8\pi^2}
\lt[-\f{1}{\ep}+\g_E-\ln(4\pi\mu^2)\rt.\nn\\
\!\!&+&\!\!\lt.\int^1_0\!\!dx\ln(M^2_N-q^2x(1-x))\rt]q^2, 
\label{pv:vac_se_01} 
\eea

\noindent where $\ep$, $\mu$ and $\g_E$ have been discussed 
previously. It is to be noted that unlike PS coupling, singularity 
in Eq.(\ref{pv:vac_se_01}) is not proportional to the mass term. 
Therefore, simple subtraction of the neutron loop contribution from 
the proton loop contribution like $\rho$-$\omega$ mixing in vacuum 
\cite{Piekarewicz92}, will remove the divergent parts. Thus the finite
$\pi$-$\eta$ mixing amplitude in vacuum is found to be

\bea
\Pi^{PV}_{vac}(q^2)\!\!&=&\!\!\f{{\rm g_\pi g_\eta}}{4\pi^2}
\lt[q^2\ln\lt(\f{M_p}{M_n}\rt)\rt. \nn \\
&+&\!\lt.\!q\sqrt{4M^2_p\!-\!q^2}\!\tan^{-1}\!
\lt(\!\f{q}{\sqrt{4M^2_p\!-\!q^2}}\!\rt)\rt.\nn \\
&-&\!\lt.\!q\sqrt{4M^2_n\!-\!q^2}\!\tan^{-1}\!
\lt(\!\f{q}{\sqrt{4M^2_n\!-\!q^2}}\!\rt)\!\rt].
\label{pv:vac_se_02}
\eea 

Similar to PS coupling, leading order contribution of the mixing 
amplitude in vacuum given by

\bea
\Pi^{PV}_{vac}({\bf q}^2)&=& - a_2{\bf q}^2, \label{pv:vac_amp} 
\eea

\noindent where $a_2=\f{{\rm g_\pi g_\eta}}{4\pi^2}\ln\lt(\f{M_p}{M_n}\rt)$.
Notice, the leading order contributions of $\pi$-$\eta$ mixing amplitudes 
in vacuum are same for both PS and PV coupling.

The three momentum dependent medium part of $\pi$-$\eta$ mixing 
self-energy reads

\bea
\Pi^{(N)}_{med}({\bf q}^2)\!\!&=&\!\!-\f{{\rm g_\pi g_\eta}}{8\pi^2M_N}
\lt[{\bf q}^2k_N\rt. \nn \\
&+&\!\!{\bf q}\lt.\lt(k^2_N-\f{{\bf q}^2}{4}\rt)
\ln\lt(\f{{\bf q}+2k_N}{{\bf q}-2k_N}\rt)\rt].
\label{pv:med_se_00}
\eea 

The leading order contribution of the medium part of the mixing 
amplitude, generated by the difference between proton and neutron 
loop contributions is 

\bea
\Pi^{PV}_{med}({\bf q}^2)\!\!&=&\!\!-b^\prime_2 {\bf q}^2,
\label{pv:med_amp}
\eea 

\noindent where 

\bea
b^\prime_2 &=&\!\!\f{{\rm g_\pi g_\eta}}{4\pi^2}
\lt(\f{k_p}{M_p}\!-\!\f{k_n}{M_n} \rt).
\eea

Notice, the density dependent mixing amplitude in PV coupling differs 
with that of PS coupling only by the term $a_1^\prime$ and this will
make difference between the medium part of the CSV potentials. 
The momentum space potential is given by

\bea
V^{NN}_{CSV}({\bf q}^2)&=&T^+_3\f{{\rm g_\pi g_\eta}}{4M^2_N}
({\bf \sigma_1}\cdot{\bf q})({\bf \sigma_2}\cdot{\bf q})\nn \\
&\times& \f{\Pi^{PV}_{\pi\eta}({\bf q}^2)}{({\bf q}^2+m^2_\pi)
({\bf q}^2+m^2_\eta)} \nn \\
&\times& \lt[1-\f{{\bf q}^2}{8M^2_N}-\f{{\bf P}^2}{2M^2_N} \rt].
\label{pv:pot_mom}
\eea

In deriveing Eq.(\ref{pv:pot_mom}) contributions from external
leg have been considered which is given within the square braket.
This contributions are same as that of PS coupling. From this 
momentum space CSV $NN$ potential one can obtain the coordinate
space potential.

\bse
\bea
V^{NN}_{vac}(r)\!\!&=&\!\!-T^+_3\f{{\rm g_\pi g_\eta}a_2}
{48\pi M^2_N}\lt[\!\f{m^5_\pi U(x_\pi)\!-\!m^5_\eta U(x_\eta)}
{m^2_\eta-m^2_\pi}\!\rt], 
\label{pv:pot_cor_vac_noff}\\
V^{NN}_{med}(r)\!\!&=&\!\!-T^+_3\f{{\rm g_\pi g_\eta}b^\prime_2}
{48\pi M^2_N}\lt[\!\f{m^5_\pi U(x_\pi)\!-\!m^5_\eta U(x_\eta)}
{m^2_\eta-m^2_\pi}\!\rt], 
\label{pv:pot_cor_med_noff}
\eea 
\ese 

The Eqs.(\ref{pv:pot_cor_vac_noff}) and (\ref{pv:pot_cor_med_noff}) 
shows the coordinate space CSV $NN$ potential without form factors.
It is to be noted that the CSV potentials in vacuum are same for both 
PS and PV couplings while density dependent parts are different. With 
form factors Eqs.(\ref{pv:pot_cor_vac_noff}) and (\ref{pv:pot_cor_med_noff}) 
reduce to

\bse
\bea
V^{NN}_{vac}(r)\!\!&=&\!\!-T^+_3\f{{\rm g_\pi g_\eta}a_2}{48\pi M^2_N}
\lt[\!\lt(\f{a_\pi m^5_\pi U(x_\pi)\!-\!a_\eta m^5_\eta U(x_\eta)}
{m^2_\eta-m^2_\pi}\rt)\rt.\nn\\
\!\!&-&\!\!\lambda\!\lt.\lt(\f{b_\pi m^5_\pi U(X_\pi)\!-
\!b_\eta m^5_\eta U(X_\eta)}{m^2_\eta-m^2_\pi}\rt) \rt], 
\label{pv:pot_cor_vac_wff}\\
V^{NN}_{med}(r)\!\!&=&\!\!-T^+_3\f{{\rm g_\pi g_\eta}b^\prime_2}
{48\pi M^2_N}\lt[\!\lt(\f{a_\pi m^5_\pi U(x_\pi)\!-
\!a_\eta m^5_\eta U(x_\eta)}{m^2_\eta-m^2_\pi}\rt)\rt.\nn\\
\!\!&-&\!\!\lambda\!\lt.\lt(\f{b_\pi m^5_\pi U(X_\pi)\!-
\!b_\eta m^5_\eta U(X_\eta)}{m^2_\eta-m^2_\pi}\rt) \rt]. 
\label{pv:pot_cor_med_wff}
\eea
\ese

\section{results}

In this section we present our numerical results. All the figures 
shows the difference between CSV $nn$ and $pp$ potentials in $^1$S$_0$
state. To obtain density dependent CSV potential we consider nuclear
matter density $\rho_B=0.148$ fm$^{-3}$ and asymmetry parameter $\a=1/3$.\\
\smallskip
\begin{figure}[htb]
\begin{center}
\resizebox{7.5cm}{5.75cm}{\includegraphics[]{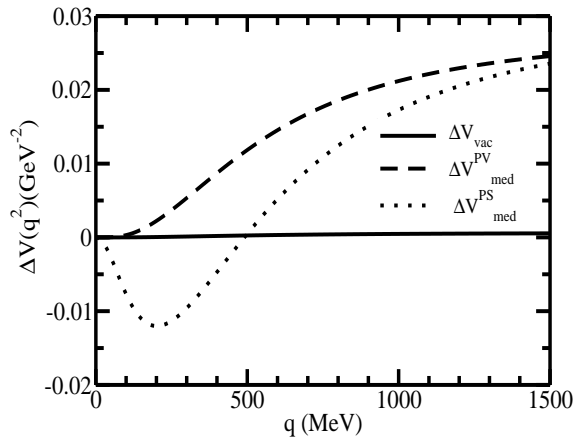}}
\caption{Difference between CSV $nn$ and $pp$ potentials
in momentum space are shown.
\label{dv_mom}}
\end{center}
\end{figure}

Fig.\ref{dv_mom} shows the difference between CSV $nn$ and $pp$ 
potentials in momentum space. The dotted and dashed curves represent 
density dependent contributions of PS and PV couplings, respectively. 
The difference in the contributions of density dependent part of CSV 
potential for these two types of coupling arises because of the term 
$a_1^{\prime}$. The vacuum contribution of CSV potentials for both PS and 
PV couplings are same. This is shown by the solid curve in 
Fig.\ref{dv_mom}.\\
\smallskip
\begin{figure}[htb]
\begin{center}
\resizebox{7.5cm}{5.75cm}{\includegraphics[]{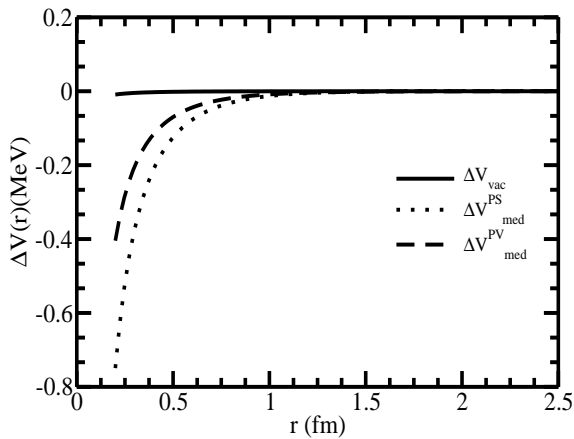}}
\caption{Difference between CSV $nn$ and $pp$ potentials 
in coordinate space without form factors.
\label{dv_cord_noff}}
\end{center}
\end{figure}

\begin{figure}[htb]
\begin{center}
\resizebox{7.5cm}{5.75cm}{\includegraphics[]{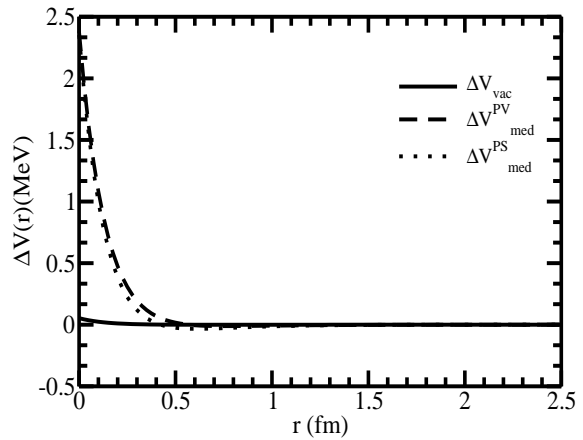}}
\caption{Difference between CSV $nn$ and $pp$ potentials
in coordinate space with form factors
\label{dv_cord_wff}}
\end{center}
\end{figure}

The CSV potential in coordinate space is presented in 
Fig.\ref{dv_cord_noff}. In this figure we show the vacuum 
and medium contribution of the CSV potential without form factors.
The same with the form factors are demonstrated in Fig.\ref{dv_cord_wff}.
Both the vacuum and medium parts contribute with the same sign. 
Note that CSV potentials change sign with the inclusion of form 
factors. Fig.\ref{dv_cord_noff} and Fig.\ref{dv_cord_wff} show 
that the medium contribution near the core region is much larger 
than the vacuum contribution. \\

The difference between  $nn$ and $pp$ scattering lengths, 
$\Dt a$, has been computed using the vacuum CSV potential
constructed here using $\pi$-$\eta$ mixing. It is found
that $\Dt a = 0.00082$ fm without form factors and with 
form factors it is $-0.0001 $ fm.\\

\section{summary and discussion}

In the present work we have constructed CS violating density dependent
two-body potential driven by the mixing of $\pi$-$\eta$  states. It is
observed that density dependent contribution is larger than the vacuum
contribution near the core region. This density dependent part might 
contribute significantly to the CSV observables. We estimate the 
contribution of $\pi$-$\eta$ mixing to the difference of $pp$ and $nn$
scattering lengths, $\Dt a$, where only the vacuum part contributes. 
Both for the density dependent and vacuum parts, we find that the role
of $\pi$-$\eta$ mixing is smaller than that of $\rho$-$\omega$ mixing 
\cite{Biswas08,Biswas10}.


\end{document}